\documentclass{emulateapj}
\usepackage{times}

\newcommand{\be}{\begin{equation}}
\newcommand{\ee}{\end{equation}}

\newcommand{\msun}{{M$_{\odot}$}}

\newcommand{\se}{s$^{-1}$ }
\newcommand{\ergs}{erg~s$^{-1}$ }
\newcommand{\ergscm}{erg~s$^{-1}$~cm$^{-2}$}
\newcommand{\degree}{$^{\circ}$} 
\newcommand{\gtsima}{$\; \buildrel > \over \sim \;$}
\newcommand{\ltsima}{$\; \buildrel < \over \sim \;$}
\newcommand{\prosima}{$\; \buildrel \propto \over \sim \;$}
\newcommand{\gsim}{\lower.5ex\hbox{\gtsima}}
\newcommand{\lsim}{\lower.5ex\hbox{\ltsima}}
\newcommand{\simgt}{\lower.5ex\hbox{\gtsima}}
\newcommand{\simlt}{\lower.5ex\hbox{\ltsima}}
\newcommand{\simpr}{\lower.5ex\hbox{\prosima}}
\newcommand{\xte}{XTE~J1650--500}

\newcommand{\nh}{N$_{\rm H}$}

\newcommand{\etal}{{et al.~}}
\newcommand{\cxo}{\textit{Chandra}}

\newcommand{\xmm}{\textit{XMM-Newton}}

\newcommand{\lx}{L$_{\rm X}$}
\newcommand{\fx}{F$_{\rm X}$}

\newcommand{\ledd}{L$_{\rm Edd}$}
\newcommand{\porb}{P$_{\rm orb}$}

\shorttitle{\cxo~detection of \xte~in quiescence}

\shortauthors{Gallo \etal}

\begin{document}

\title{{\it CHANDRA} detection of \xte~in
quiescence and the minimum luminosity of black hole X-ray binaries}

\author{Elena Gallo\altaffilmark{1,2},
 Jeroen Homan\altaffilmark{3}, Peter G. Jonker\altaffilmark{4,5}, John A. 
Tomsick\altaffilmark{6}}
\altaffiltext{1}{Physics Department, Broida Hall, University of
California Santa Barbara, CA 93106} 
\altaffiltext{2}{Chandra Fellow}
\altaffiltext{3}{Kavli Institute for Astrophysics and Space Research,
Massachusetts Institute of Technology, 77 Massachusetts Av., Cambridge, MA
02139}
\altaffiltext{4}{SRON, Netherlands Institute for Space Research, Sorbonnelaan 2, 3584~CA, Utrecht, NL
}
\altaffiltext{5}{Harvard--Smithsonian  Center for Astrophysics, 60 Garden
Street, Cambridge, MA~02138 }
\altaffiltext{6}{Space Sciences Laboratory, 7 Gauss Way, University of
California Berkeley, CA 94720}

\begin{abstract}
The Galactic black hole X-ray binary \xte~entered a quiescent regime following
the decline from the 2001--2002 outburst that led to its discovery. Here we
report on the first detection of its quiescent counterpart in a 36 ks
observation taken in 2007 July with the \cxo~{\it X-ray Observatory}. The
inferred 0.5--10 keV unabsorbed flux is in the range $2.5$--$5.0\times
10^{-15} $ \ergscm. 
Notwithstanding large distance uncertainties, the measured luminosity
is comparable to that of the faintest detected black hole X-ray binaries, all
having orbital periods close to the expected bifurcation period between $j$-
and $n$-driven low-mass X-ray binaries. This suggests that a few $10^{30}$
\ergs~might be a limiting luminosity value for quiescent black holes.
\end{abstract}


\section{Introduction}
\label{sec:intro}

Black hole (BH) X-ray transients -- close binary systems in which a low-mass
donor transfers mass via Roche-lobe overflow onto a black hole accretor --
spend most of their lifetimes in a low-luminosity state, where the boundary
between `quiescence' and more active regime can be set around $10^{33.5}$ erg
\se, corresponding to a few $10^{-6}$ times the Eddington luminosity (\ledd)
for a 10~\msun~object (e.g. McClintock \& Remillard 2006).
Broadly speaking, the low Eddington ratios can be due to low radiative
efficiency or low net accretion rate in the inner regions, or a combination of
the two (see e.g. Narayan 2005, and references therein). First explored in
their basic ideas by Ichimaru (1977), and Rees \etal (1982), stable,
radiatively inefficient accretion flow models were later formalized and gained
vast popularity owing to the works of Narayan \& Yi (1994, 1995), and Abramowicz \etal
(1995). Since the mid 1990s, they have been widely employed to account for the
broadband (radio/optical/UV/X-ray) spectra of low-luminosity BH candidates,
such as A0620--00, GS 1124--68 and V404 Cyg (Narayan \etal 1996, 1997a), as
well as the Galactic center source Sgr A$^{\star}$ (Narayan \etal 1995).
The increased sensitivity of \cxo~and
\xmm~with respect to previous X-ray telescopes has eventually
permitted detailed X-ray studies of quiescent Galactic BHs down to Eddington
ratios as low as a few $10^{-9}$ (Garcia \etal 2001; Kong \etal 2002; Hameury
\etal 2003; Tomsick
\etal 2003; McClintock \etal 2003; Gallo \etal 2006; Homan \etal 2006; Corbel
\etal 2006; Bradley \etal 2007).  
In the framework of
advection-dominated accretion flow models (ADAF; Narayan \& Yi 1994, 1995),
the observed luminosity difference between quiescent BHs and quiescent neutron
stars -- the former being fainter by about one order of magnitude at
comparable orbital periods -- has been interpreted as evidence for the
existence for an event horizon in BHs (Narayan \etal 1997b; Menou \etal 1999;
Garcia \etal 2001; Narayan \& McClintock 2008).
At the same time, recent studies at lower frequencies, in the radio and
mid-infrared bands, suggest that BHs and neutron stars may channel different
fractions of the total accretion power into relativistic outflows, with a
substantially reduced jet contribution in quiescent neutron stars
with respect to BHs (Fender \etal 2003; Gallo \etal 2006, 2007; Migliari \etal
2006; K{\"o}erding \etal 2007).

Of the 40 candidate BH X-ray binaries\footnote{Additional likely BH candidates
have been since discovered, including: XTE J1817--330, Swift J1753.5--0127,
IGR J17091--3624, IGR J17098--3628, IGR J17497--2821 and IGR J18539+0727.} 
listed by Remillard \& McClintock (2006), 15 have sensitive measurements/upper
limits on their quiescent X-ray luminosities.
In this Letter, we report on a 36 ks observation of the quiescent BH XTE
J1650--500 performed in 2007 July with the \cxo~{\it X-ray Observatory}, and
briefly discuss it in the context of quiescent BH X-ray binaries and how the
advent of high sensitivity/resolution X-ray telescopes has improved our
understanding of such systems.

The Galactic X-ray binary system \xte~entered a quiescent regime following the
2001--2002 outburst that led to its discovery with the {\it Rossi X-ray Timing
Explorer} (Remillard 2001).  Observations conducted with \xmm~and {\it
BeppoSAX} in late 2001, right after the outburst peak, revealed a broad,
asymmetric Fe K$\alpha$ emission line, interpreted as due to irradiation of
the inner accretion disk around a rapidly spinning Kerr BH (Miller \etal 2002;
Miniutti \etal 2004; see Done \& Gierlinski 2006 for a different
interpretation). The prolonged quiescent regime has allowed for the derivation
of the system orbital period and optical mass function: P$_{\rm orb}$=7.7 hr,
and $f(M)$=2.73$\pm$0.56 \msun, respectively (Orosz \etal 2004).
The mass of the BH in \xte~is highly uncertain. The amplitude of
the phased $R$-band light curve results in a lower bound to the orbital
inclination $i>50$\degree$\pm$3\degree, which, in the limiting case of no disk
contribution, implies in an upper limit of 7.3 \msun~to the
accretor mass (Orosz \etal 2004). However --caveat the poor signal-to-noise
of the adopted stellar templates-- these authors find that the accretion
disk contribution in the $R$ band can be as high as $\simgt$80$\%$.
This results in an orbital inclination $i=70$\degree~or higher, that is a mass
of only 4 \msun~for the BH.

The non-detection of \xte~with \xmm, in 2005 March, yielded an upper limit on
its quiescent unabsorbed flux of 1.1--1.3$\times 10^{-14}$ \ergscm~(0.5--10
keV), depending on the assumed spectral model (Homan \etal 2006).  At a (by no
means certain\footnote{Similarly to its mass, the distance to \xte~is also
uncertain: the quoted value was estimated by Homan \etal (2006) based on the
system Eddington-scaled X-ray luminosity at the soft-to-hard X-ray state
transition, following Maccarone \& Coppi (2003). As such, it suffers from
uncertainties of at least a factor 2.}) distance of 2.6 kpc, this corresponds
to a X-ray luminosity \lx$\simlt 10^{31}$ \ergs, comparable to the lowest
luminosities inferred for quiescent BH X-ray binaries (i.e. GRO J1655--40, GRO
J0422+32, XTE J1118+480, A0620--00 and GS 2000+25).

\def\errtwo#1#2#3{$#1^{+#2}_{-#3}$}

\begin{deluxetable*}{crcccc} 
\setlength{\tabcolsep}{0.07in} 
\tabletypesize{\scriptsize} 
\tablewidth{0pt} 
\tablecaption{\cxo~detected sources in the \xte~field.\label{tab:list}}
\tablehead{  \colhead{Source}
	   & \colhead{Name}
           & \colhead{R.A.} 
           & \colhead{Decl.}
          & \colhead{Count rate} 
	   & \colhead{Notes} 
\\
           \colhead{(1)}
           & \colhead{(2)} 
           & \colhead{(3)} 
	   & \colhead{(4)} 
           & \colhead{(5)}
	   & \colhead{(6)}
          } 
\startdata 
1  & CXOU J165007.6-495623 & 16 50 07.69 & -49 56 23.8 & 1.9 (0.7) \\
2  & CXOU J165000.2-495723 & 16 50 00.21 & -49 57 23.4 & 1.9 (0.7) \\
3  & {\bf XTE J1650-500}   & 16 50 00.92 & -49 57 44.1 & 2.0 (0.7) \\
4  & CXOU J165006.3-495814 & 16 50 06.37 & -49 58 14.3 & 2.0 (0.7) \\
5  & CXOU J165013.1-495709 & 16 50 13.14 & -49 57 09.1 & 2.0 (0.7) \\
6  & CXOU J165005.4-495427 & 16 50 05.48 & -49 54 27.6 & 2.1 (0.8) \\
7  & CXOU J165002.5-495333 & 16 50 02.56 & -49 53 34.0 & 2.2 (0.8) \\
8  & CXOU J164943.9-495901 & 16 49 43.97 & -49 59 01.3 & 2.2 (0.8) \\
9  & CXOU J165012.1-495715 & 16 50 12.12 & -49 57 15.2 & 2.4 (0.8) \\
10 & CXOU J165000.3-495655 & 16 50 00.38 & -49 56 55.9 & 2.5 (0.8) \\
11 & CXOU J165006.0-495455 & 16 50 06.05 & -49 54 55.1 & 2.5 (0.8) & USNO B1  \\
12 & CXOU J165021.3-495632 & 16 50 21.38 & -49 56 32.0 & 3.0 (0.9) \\
13 & CXOU J164950.3-495931 & 16 49 50.30 & -49 59 31.0 & 3.1 (0.9) \\
14 & CXOU J165006.3-495743 & 16 50 06.35 & -49 57 44.0 & 3.6 (1.0) \\
15 & CXOU J165009.1-495442 & 16 50 09.10 & -49 54 42.9 & 4.2 (1.1) & USNO B1  \\
16 & CXOU J164959.7-495518 & 16 49 59.77 & -49 55 18.7 & 4.2 (1.1) \\
17 & CXOU J164958.5-495614 & 16 49 58.54 & -49 56 14.2 & 4.8 (1.2) \\
18 & CXOU J164951.8-495653 & 16 49 51.87 & -49 56 53.8 & 5.2 (1.2) \\
19 & CXOU J165013.9-495726 & 16 50 13.94 & -49 57 26.7 & 5.3 (1.2) \\
20 & CXOU J164947.8-500119 & 16 49 47.82 & -50 01 19.9 & 5.4 (1.2) \\
21 & CXOU J164955.5-495705 & 16 49 55.50 & -49 57 05.6 & 5.6 (1.2) \\
22 & CXOU J165005.2-495622 & 16 50 05.27 & -49 56 22.6 & 5.9 (1.3) \\
23 & CXOU J164943.2-495450 & 16 49 43.28 & -49 54 50.6 & 7.2 (1.4) \\
24 & CXOU J165005.1-495624 & 16 50 05.13 & -49 56 24.3 & 7.2 (1.4) & USNO B1 \\
25 & CXOU J164956.0-495711 & 16 49 56.04 & -49 57 11.2 & 8.0 (1.5) \\
26 & CXOU J164953.5-495747 & 16 49 53.59 & -49 57 47.6 & 8.6 (1.5) \\
27 & CXOU J164948.8-495509 & 16 49 48.82 & -49 55 09.8 & 9.4 (1.6) \\
28 & CXOU J164948.7-495936 & 16 49 48.72 & -49 59 36.2 &11.5 (1.8) \\  
\enddata \tablecomments{(1) Target numeration (2) Source name following the
\cxo~convention (3) Units of right ascension (equinox J2000.0) are
hours, minutes, seconds; (4) Units of declination (equinox J2000.0) are degrees,
arcminutes and arcseconds; (5) Net count rate, in units of $10^{-4}$ cps, as
measured by {\tt wavdetect}, with errors given in parenthesis, over
the energy interval 0.3--7.0 keV. Notice that {\tt wavdetect} is designed to
be used as a detection algorithm, and only secondarily as a source flux
measurement tool. Count rates are generally reliable, but can be slightly
under-estimated for very low number of counts. } 
\end{deluxetable*} 

\section{Observation and data analysis}
\label{sec:data}

The field of view of \xte~was observed with with the Advanced CCD Imaging
Spectrometer (ACIS) detector on board \cxo~on 2007 June 30 at 23:46 UT for
approximately 36 ks. The target was placed on the back-side illuminated S3
chip in order to take advantage of the detector sensitivity to low-energy
X-rays. The data were telemetered in very faint (VF) mode with an upper energy
cut-off at 13 keV, in order to minimize the chances of saturation.
We have reprocessed and analyzed the data using the \cxo~Interactive Analysis
Observation (CIAO) software version 3.4.1.1. The
level 1 event lists were first cleaned following the standard threads to
reduce the ACIS particle background for VF mode observations, including only
ASCA grades 0,2,3,4 and 6. Further analysis was restricted between 0.3--7.0 keV in
order to avoid calibration uncertainties at low energies and to limit
background contaminations at high energies.
As \cxo~is known to encounter periods of high background, which especially
affect the S1 and S3 chips, we first checked for background flares but found
none, resulting in a net exposure of 35.65 ks. We applied a wavelet
detection algorithm, using CIAO {\tt wavdetect} with a sensitivity threshold that
corresponds to a $3.8\times10^{-6}$ chance of detecting a spurious source per point
spread function (PSF) element if the local background is uniformly distributed.
We employed the default `Mexican Hat' wavelet, with scales increasing by
a factor of $\sqrt{2}$ between 1--16 pixels  
on a full resolution circular region of 512 pixel radius centered on the nominal position of the target
(restricting the circle to the S3 chip).
Table 1 lists the 28 sources detected by the
algorithm, in order of increasing count rate (the positions were obtained after registering the
\cxo~image to the USNO B1 catalog, see below).  
Three of the detected X-ray sources have optical counterparts with positions
listed in the US Naval Observatory (USNO) B1 catalog (Monet \etal 2003), which has an absolute
positional error of 0\arcsec.20. 
The mean USNO to \cxo~shifts in right ascension $\alpha$ and declination
$\delta$ are $0\arcsec.06\pm0\arcsec.31$ and
$0\arcsec.08\pm0\arcsec.31$, respectively, where the uncertainties account for
the \cxo~statistical errors as well as the USNO B1 positional uncertainty.
We thus registered the \cxo~image by applying the above astrometric
corrections, and ran {\tt wavdetect} again on the registered image to obtain refined
position of the X-ray sources. 

\begin{figure}[t!]
\begin{center}
\includegraphics[angle=0,scale=.425]{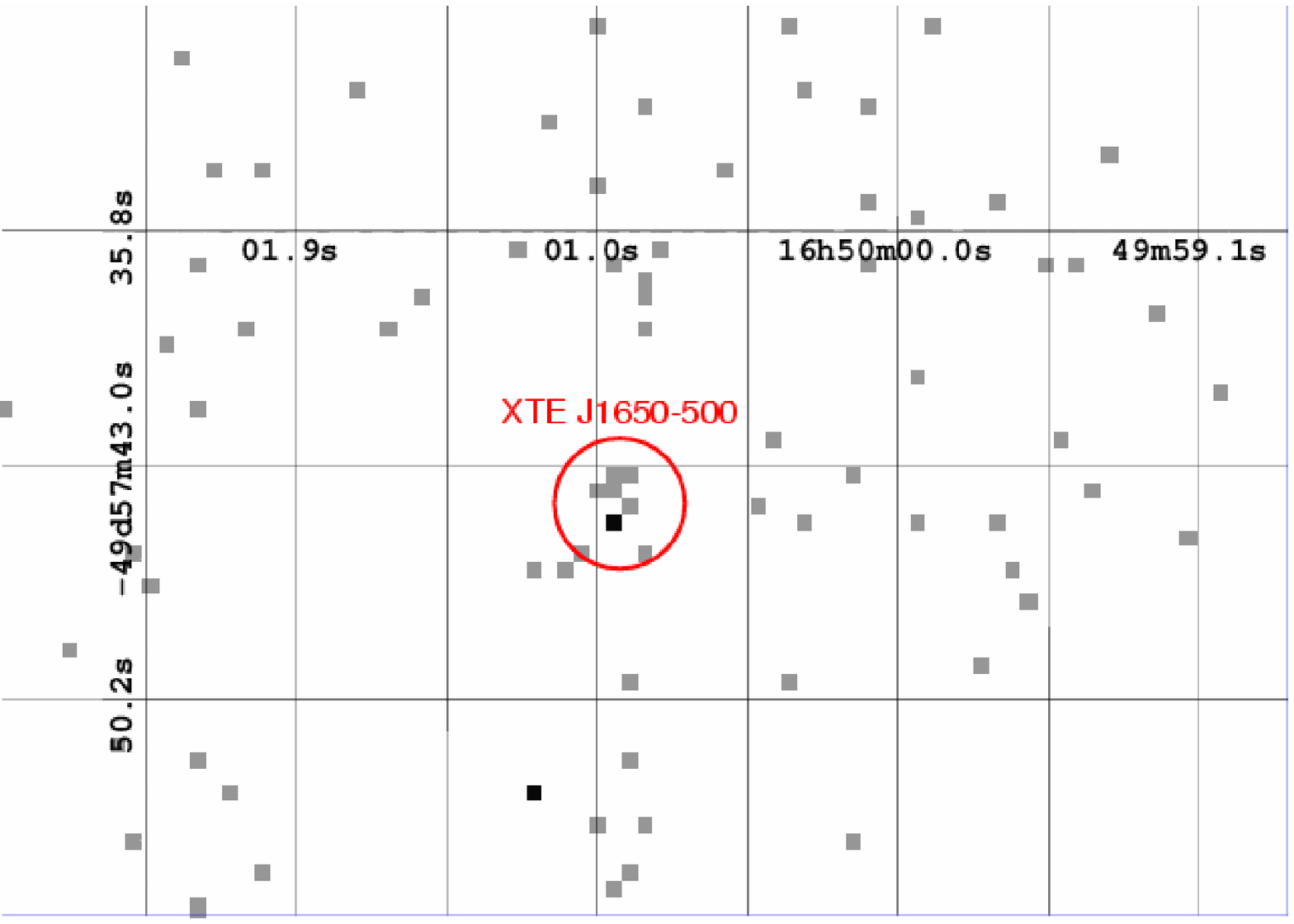}
\caption{\cxo~image of the field around \xte~(ACIS-S3, 0.3--7 keV). 
The red circle -- centered on the source position as identified by the CIAO
{\tt wavdetect} -- marks the count extraction region.
The astrometry has been improved by
cross-matching the \cxo~image to the USNO B1 catalog; the revised
\cxo~position of \xte~is: R.A.:16$^{\rm h}$:50$^{\rm m}$:00$^{\rm s}$.92,
decl.:$-$49\arcdeg:57\arcmin:44\arcsec.1 (J2000.0), with an uncertainty
of 0\arcsec.31.  For reference, 1 pixel corresponds to 0\arcsec.492.  North is
at the top and East is on the left on this image.  \label{fig:fov}}
\end{center}
\end{figure}

The position of source 3, shown in Figure~\ref{fig:fov}, is consistent with the X-ray position
of XTE J1650--500 reported by Tomsick \etal (2004) based on \cxo~observations
taken during a higher luminosity state:
after the astrometric correction, the revised 
position of \xte~is: $\alpha$=16$^{\rm h}$:50$^{\rm
m}$:00$^{\rm s}$.92, $\delta$=$-$49\arcdeg:57\arcmin:44\arcsec.1 (equinox J2000.0), 
with an uncertainty of 0\arcsec.31.
In order to determine the optimal extraction region for the source counts, we
generated the ACIS PSF at 1.5 keV at the position of
\xte, and (arbitrarily) normalized it to 10$^4$ counts. This allows for an
evaluation of R$_{\rm 95\%}$, the radius which encircles 95$\%$ of the energy;
At this angular off-set (0\arcmin.302), R$_{\rm 95\%}$=2\arcsec~at 1.5 keV. A
total of 9 counts are measured within a circle with 2\arcsec~radius centered on
the position of \xte~given above. The inferred number of {\it net} counts is
$8.2\pm 2.9$.
In order to quantify the significance of the detection, we compared the total
number of counts within a 2\arcsec\ radius circular aperture, and within a
background annulus with inner and outer radii $R_{\rm in}=$5\arcsec\ and
$R_{\rm out}$=25\arcsec, both centered on the target position.  We found 9 
counts within the 2\arcsec\ radius aperture, and 122 counts within the
background annulus. The Poisson probability of obtaining 9 counts or more when
0.8 are expected is $\sim 2\times 10^{-7}$, implying a detection at the
99.99998$\%$ confidence level (e.g. Weisskopf \etal 2007).
The low number of counts detected from \xte~does not allow for a proper
spectral modeling.  
We estimated an approximate flux using webPIMMS\footnote{http://heasarc.nasa.gov/Tools/w3pimms.html}.  
Assuming an absorbed power law model with photon index $\Gamma=1.7-2$ (typical
for quiescent BHs, e.g. Corbel \etal 2006) and equivalent hydrogen column
\nh$=6.7\times 10^{21}$ cm$^{-2}$ (Tomsick \etal 2004), the measured net count
rate of $2.3\pm0.8\times 10^{-4}$ cps results in an unabsorbed flux
\fx$=3.7\pm 1.3 \times 10^{-15}$ \ergscm~over the energy interval 0.5--10.0
keV (the quoted flux uncertainty accounts for both counting statistics and the
different assumed spectral shapes, being dominated by the former).
This yields a luminosity \lx$\simeq 3\pm 1\times 10^{30}~(\rm
D/2.6~kpc)^2$
\ergs~over the same energy interval, well below the upper limit from
\xmm~(Homan \etal 2006). We
wish to stress once again the large uncertainty in the distance, whose
estimation method in turn relies on the uncertain BH mass value (see Appendix
in Homan \etal 2006).
For a maximum BH mass of 7.3 \msun, the inferred luminosity corresponds to a minimum
Eddington-scaled luminosity \lx$\simgt 3\times 10^{-9}$
\ledd~(Orosz \etal 2004). 

\section{Discussion} 

\begin{figure*}[t!]
\begin{center}
\includegraphics[angle=0,scale=.4]{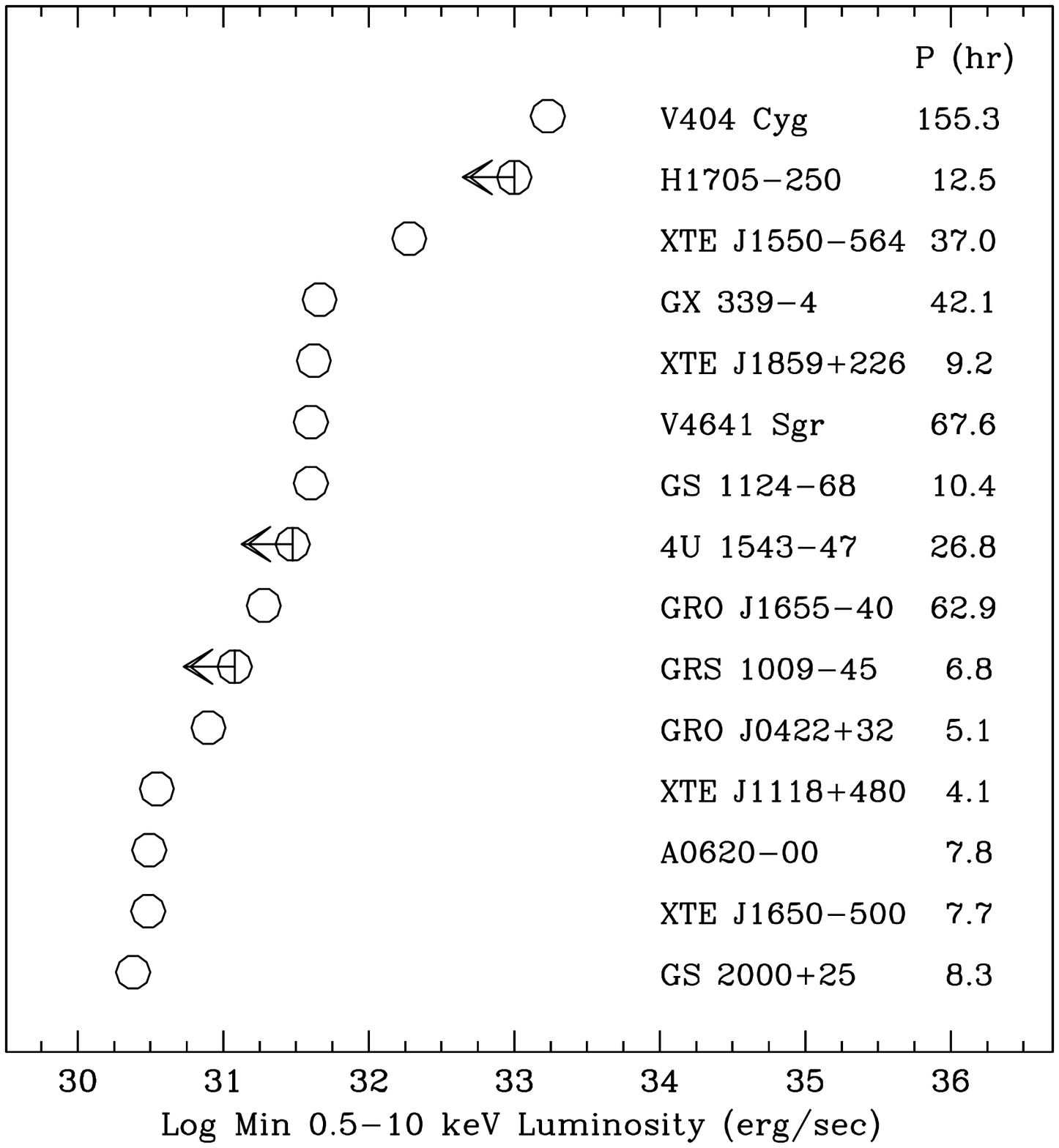}
\hspace{0.75cm}\includegraphics[angle=0,scale=.4]{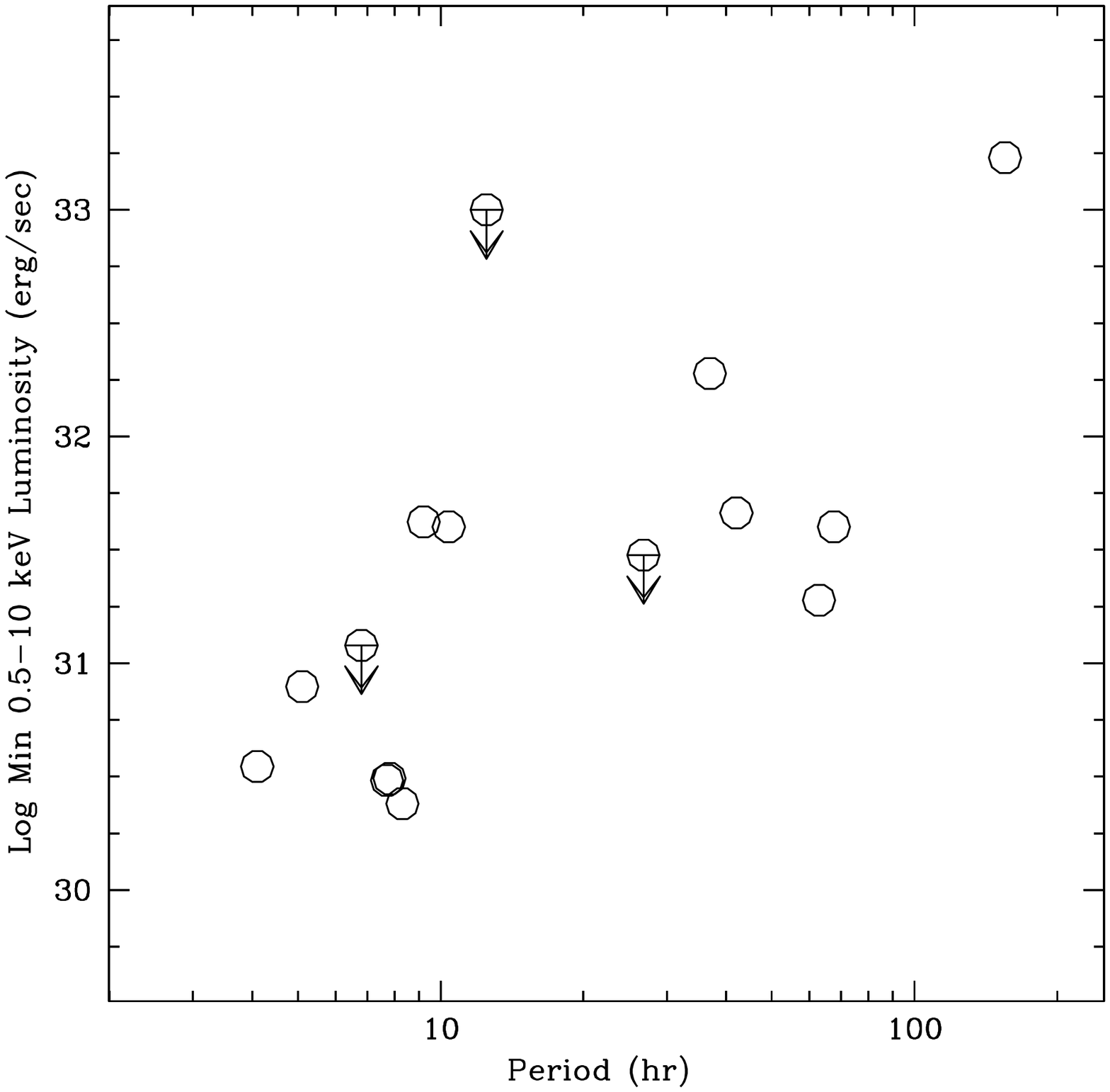}
\caption{
{\it Left}:  Quiescent X-ray luminosities/upper limits for 15 BH X-ray binaries with
sensitive observations (down to a minimum threshold of $\sim 10^{33.5}$
\ergs). 
{\it Right}: The measured luminosities are plotted against the systems'
orbital periods. The faintest systems (GS~2000+25, A0620--00 and XTE J1650--500)
have orbital periods close to 10 hr, the expected bifurcation period between
$j$- and $n$-driven low-mass X-ray binaries according to Menou \etal (1999). \label{fig:qbh}}
\end{center}
\end{figure*}

As illustrated in Figure~\ref{fig:qbh} (updated from Tomsick \etal 2003 after
Corbel \etal 2006, Gallo \etal 2006 and Homan \etal 2006), the quiescent X-ray
luminosity of \xte~as measured by \cxo~sits in the range of values inferred
for systems with similar orbital periods. Out of 15 candidate BH X-ray binary
systems with sensitive observations while in the quiescent regime, 12 have now
been detected in X-rays.  For those 12, the quiescent luminosities range
between a few $10^{30}$ and $10^{33}$ \ergs. The nearest BH, A0620--00, has
been steadily emitting at $\simeq 2-3\times 10^{30}$ erg \se at least for the
past 5 years (Kong \etal 2002; Gallo \etal 2006); this is approximately the
same luminosity level as XTE J1650--500 (this work), XTE J1118+480 (McClintock
\etal 2003) and GS~2000+25 (Garcia \etal 2001), suggesting that this might be
some kind of limiting value.

Indeed, for low-mass X-ray binaries one can make use of binary evolution
theory, combined with a given accretion flow solution, to predict a relation
between the minimum quiescent luminosity and the system orbital period
\porb~(see e.g. Menou \etal 1999; Lasota 2000). As an example, Figure 4 of
Menou \etal (1999) illustrates how the predicted luminosity of quiescent BHs
powered by ADAFs depends on the ratio between the outer mass transfer rate and
the ADAF accretion rate.  The lower band, which roughly reproduces the
observed luminosities of 3 representative systems spanning the whole range of
detected luminosities (A0620--00, GRO J1655--40 and V404 Cyg), corresponds to
$\sim$1/3 of the mass transferred being accreted via the ADAF, whereas the
remaining 2/3 would be either accumulated in an outer thin disk or lost to an
outflow.  Most importantly, independently of the actual solution for the
accretion flow in quiescence, the existence of a minimum luminosity in
low-mass X-ray binaries stems directly from the existence of a bifurcation
period, P$_{\rm bif}$, below which the mass transfer rate is driven by
gravitational wave radiation ($j$-driven systems), and above which it is
dominated by the nuclear evolution of the secondary star ($n$-driven
systems). Specifically, the outer mass transfer rate $\dot{M}_{\rm T}$
increases with decreasing orbital period below P$_{\rm bif}$, while 
increases with increasing orbital period above P$_{\rm bif}$ (the same applies
to quiescent neutron star X-ray binaries, although with higher normalization).
For a wide range of donor masses, Menou \etal find P$_{\rm bif}\simeq 10$ hr.
{\it As long as the luminosity expected from a given accretion flow model
scales with a positive power of $\dot{M}_{\rm T}$, systems with orbital
periods close to the bifurcation period should display the lowest quiescent
luminosities}. Caveat the large distance uncertainties which affect most BH
X-ray binaries (e.g. Jonker \& Nelemans 2004), this is indeed observed, as
illustrated in the right panel of Figure~\ref{fig:qbh}.
%

Deep observations of short ($\simlt$5 hr), as well as long ($\simgt$200 hr) 
orbital period systems will hopefully add to the
slim statistics in support of this argument. 
We wish to point out that, given the distance/absorption column to the 3 BH
systems with upper limits (H1705--250, GRS1009--45 and 4U~1543--47; see
Figure~\ref{fig:qbh}), several tens of ks of integration with
\cxo~would be needed in order to detect them, should their luminosity
be close to that of the faintest objects.
Quiescent ultra-compact X-ray binaries (with orbital periods
\porb$\simlt$1 hr) could be primary targets to test this conclusion at the
short \porb~end. However, the argument has been made that the mass-transfer
rate evolution has not been studied in any detail for systems with such very
low mass ratios. Extreme mass ratios could result in a drastically reduced
$\dot{M}_{\rm T}$, making the direct comparison between
quiescent ultra-compact and longer period~binaries somewhat dubious (see
Jonker \etal 2006, 2007, and related discussion in Lasota 2007). For long
orbital period systems, the ideal testbed would be obviously provided by the
power off of the as yet super-luminous GRS~1915+105, with an orbital period of
over 800 hr. Similarly, the newly discovered BH candidate Swift J1753.5--0127,
with an inferred orbital period of 3.2 hr (Zurita \etal 2008), represents the
ideal short period target.
\acknowledgments E.G. is funded by NASA through \cxo~Postdoctoral
Fellowship grant number PF5-60037, issued by the {\it Chandra X-ray
Observatory} Center, which is operated by the Smithsonian
Astrophysical Observatory for NASA under contract NAS8-03060. Support
for this work was provided through \cxo~Award Number G07-8047A issued
by the {\it Chandra X-ray Observatory} Center. We wish to thank the
referee for a prompt and thoughtful report.

\end{document}